# Latency Optimization for Resource Allocation in Cloud Computing System

Masoud Nosrati[1(✉)], Abdolah Chalechale[2], and Ronak Karimi[1]

[1] Kermanshah Branch, Islamic Azad University, Kermanshah, Iran
minibigs_m@yahoo.co.uk
[2] Department of Computer Engineering, Razi University, Kermanshah, Iran

**Abstract.** Recent studies in different fields of science caused emergence of needs for high performance computing systems like Cloud. A critical issue in design and implementation of such systems is resource allocation which is directly affected by internal and external factors like the number of nodes, geographical distance and communication latencies. Many optimizations took place in resource allocation methods in order to achieve better performance by concentrating on computing, network and energy resources. Communication latencies as a limitation of network resources have always been playing an important role in parallel processing (especially in fine-grained programs). In this paper, we are going to have a survey on the resource allocation issue in Cloud and then do an optimization on common resource allocation method based on the latencies of communications. Due to it, we added a table to Resource Agent (entity that allocates resources to the applicants) to hold the history of previous allocations. Then, a probability matrix was constructed for allocation of resources partially based on the history of latencies. Response time was considered as a metric for evaluation of proposed method. Results indicated the better response time, especially by increasing the number of tasks. Besides, the proposed method is inherently capable for detecting the unavailable resources through measuring the communication latencies. It assists other issues in cloud systems like migration, resource replication and fault –tolerance.

## 1 Introduction

The distributed computers emerged to tie together the power of large number of resources distributed across a network [1]. The requirements of each user are shared on the network through a proper communication channel. It helps to utilize the capabilities of the whole of system for all users. Generally, High Performance Computing Systems (HPC) are defined as a collection of single coherent systems that are interconnected through a high speed network, to provide facility of high performance computing [2]. Many applications of High Performance Computing systems such as industrial [3], educational [4], medical [5] and commercial [6] came to existence after provision of hardware infrastructures. HPCs are preferred for the following reasons [7]:

- The nature of distributed applications is based on the network connections.
- Parallelism is provided in HPC by executing parallel grains on different machines.
- Higher reliability rather than single systems.

Previous studies in this area, categorizes the distributed systems to 3 types: Cluster, Grid and Cloud. In classic texts, Cluster is known as a distributed system with homogeneous nodes and Grid with heterogeneous ones [8][9]. But, in recent researches, there are Clusters with heterogeneous nodes implemented. It shows that homogeneity can't be a good metric for classification. Due to it, [7] categorizes the distributed systems by their resource allocation features as the fig.1. In this study, we will get into the resource allocation in Cloud systems.

Modeling of a Cloud system is not an easy errand to run. It is complicated because of wide range of different factors influencing the systems, such as: number of machines, types of applications, processing load and other important factors which can affect the system. Type of services is also a critical point. It can be Software as a Service (SaaS), Platform as a Service (PaaS) and Infrastructure as a Service (IaaS). Fig.2 shows some instances of these services. Another important notion is the issue of migration. Migration lets the system to achieve better performance, fault-tolerance, system management and load distribution. So, it is always considered in design and modeling of all Clouds. Quality of Service is other challenge in Cloud systems. QoS can be set in run time. This feature is called "Dynamic QoS Negotiation" [10]. DQoSN is implemented by a special entity or through self-algorithms [10].

The challenges that were talked can clearly indicate the delicacy and toughness of resource allocation in Cloud systems. Accordingly, [11] classifies the recent studies in Cloud resource allocation in 3 types:

- Researches focusing on processing resources like [12] and [13].
- Researches focusing on network resources like [14] and [15].
- Researches focusing on power and energy resources like [16] and [17].

Also, [11] states that challenges are external or internal. External challenges include regulative and geographical challenges (it is about the geographical location, and regulative and security issues of data caused by distribution) and charging model issues (it is about the charges for customers to utilize the Cloud). In other hand, internal challenges include the data locality: combining compute and data management; reliability of network resources inside a data center; and Software Defined Networking design challenges inside the data centers. SDN is a networking paradigm that separates the forwarding plane from software control. Details are mentioned in [11].

One of the important issues in real world implementation of distributed systems is about the power and energy resources optimization. Different strategies are introduced in order to decrease the energy consumption. Most of them try to aggregate the resources on smaller number of servers, in order to shut down the non-busy ones. Resource aggregation has a trade-off with the performance of the system. Also, it will cause to have a bottle neck on the I/O of the running servers. Also, it may affect the Quality of Service of system. Average response time will be increased, respectively.

As the conclusion of this section, it should be restated that in every resource allocation method, all the aspects of processing capabilities, network resources and energy consumption must be considered. Regarding the trade-off among these factors, the best configuration should be found and set. Besides, other issues like the portability and fault-tolerance should be paid attention, respectively. Recent studies go through these issues separately and many approaches for optimization of recourse management are offered by them.

In the rest of this paper, we will have a brief look at the resource allocation and the strategies that Resource Agent (the entity that allocates the resources to applicants) utilizes to choose the best resource for best applicant. After pointing out the standard method of resource allocation, we offer our contribution based on the optimization of latencies between the resource and applicant. Results of simulation of proposed method shows the better performance of this method rather than the standard resource allocation approach. Finally it the end of paper, we will have a discussion and conclusion on the features of proposed method.

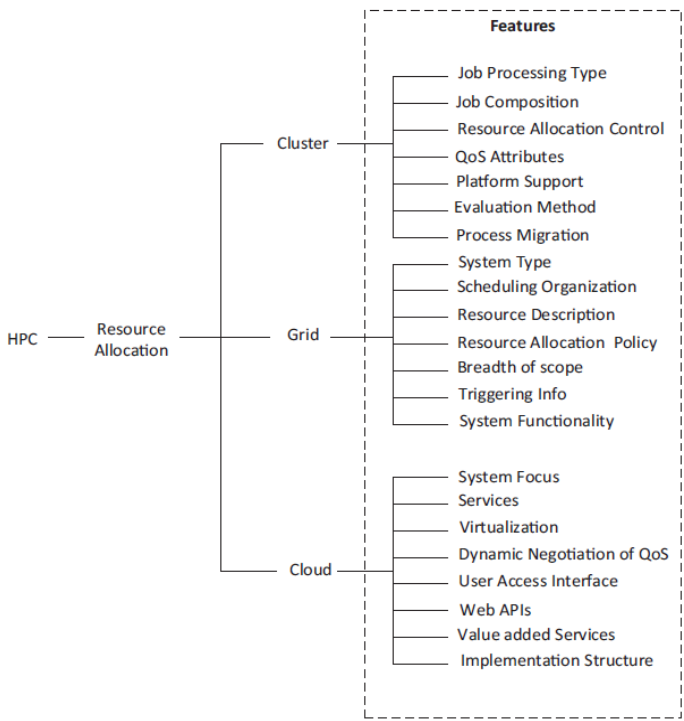

**Fig. 1.** HPC systems categories and attributes [7]

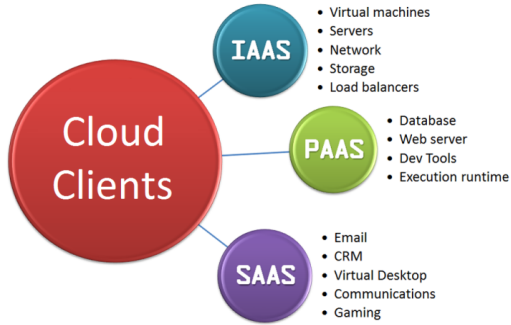

**Fig. 2.** Instances of Cloud services
Source: http://ohsweb.ohiohistory.org/ohioerc/?page_id=187

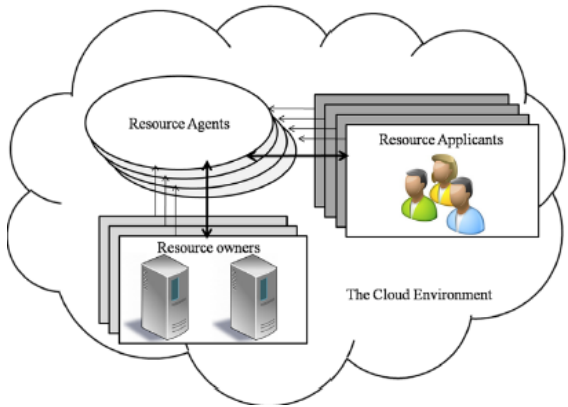

**Fig. 3.** Cloud resources allocation environment [21]

## 2      Common Resource Allocation Method

In this section, a common resource allocation method is talked that many researches like [18], [19] and [20] utilized it for their resource allocation optimization. In this method, a resource agent is considered as the entity that performs the resource allocation. As it is shown in fig.3, both resource owners and resource applicants send their costs to ResAg. General policy of ResAg is to sort the requests and resources and totally allocate the applicants with highest budget to the resource with lowest price [21].

Let $U = \{u_1, u_2, u_3, \ldots, u_m\}$ be the set composed of m Resource applicants, each task of resource applicant $u_i$ is $t_i$, so the task set of $U$ can be described as $T = \{t_1, t_2, t_3, \ldots, t_m\}$. And $t_i$ has four attributes $t_i = \{tid_i, l_i, b_i, d_i\}$, where $tid_i$ is the *ith* task's identify, $l_i$ is the $i^{th}$ task's length, $b_i$ is the $i^{th}$ task's budget, and $d_i$ is the deadline of the task.

Let $O = \{o_1, o_2, o_3, ..., o_n\}$ be the set composed of n resource owners, each resource of resource owner $o_j$ is $r_j$, so the resource set of $O$ can be described as $R = \{r_1, r_2, r_3, ..., r_n\}$. And $r_j$ has five attributes $r_j = \{rid_j, cpu_j, st_j, lp_j, hp_j\}$, where $rid_j$ is the $j^{th}$ media resource's identify, $cpu_j$ is the $j^{th}$ media resource's computing ability of solving the task, $st_j$ is the start time to deal with a new task (i.e. the current workload of resource $r_j$), $lp_j$ is the $j^{th}$ media resource's lowest price, and $hp_j$ is the $j^{th}$ media resource's highest price.

The media resources allocation probability matrix is shown as $P$ which each $p_{ij}$ is the probability of resource $j$ to be allocated to applicant $i$:

$$P = \begin{bmatrix} p_{11} & p_{12} & \cdots & p_{1n} \\ p_{21} & p_{22} & \cdots & p_{2n} \\ \vdots & \vdots & \vdots & \vdots \\ p_{m1} & p_{m2} & \cdots & p_{mn} \end{bmatrix} \quad \text{s.t.} \quad 0 \leq p_{ij} \leq 1 \;,\; \sum_{i=1}^{m} p_{ij} = 1 \;,\; \sum_{j=1}^{n} p_{ij} = 1$$

Before any decision about resource allocation, the budget of applicant and the price of resource should be calculated and submitted to ResAg. It is important to consider the following point all the time:

Resource must be capable enough to process the request of applicant in the deadline: $d_i - st_j - \frac{l_i}{cpu_j} \geq 0$

The price of resource must be less than or equal to the applicant: $b_i/l_i \geq lp_j$

Budget of applicant must be at least equal to the average of the price of remaining resources. $(\overline{lp} = \left(\frac{1}{n}\right) \sum_{j=1}^{n} lp_j)$

Budget of applicant is calculated from (1), where $bid_i^{resource}$ has inverse relationship with the number of remaining resources. It means, when the number of unallocated resources is decreasing, proposed budget of applicant for the resource will be increased, and vice versa. Other impressing factor that is Average Remaining Time, that [22] calculated it as (2). Total budget based on the (3) is the sum of both (1) and (2) with the weights of $\alpha'$ and $\beta'$. Weights might be changed based on the policies of the system.

$$bid_i^{resource}(t) = \overline{lp} + \left(\frac{b_i}{l_i} - \overline{lp}\right)\left(1 - \frac{n_i^t}{n_i^{max}}\right)^{\frac{1}{\alpha}} \tag{1}$$

$$\overline{rt}_i(t) = \sum_{j=1}^{n} \frac{(rt_{ij}(t)\omega_{ij})}{n_i^{max}} \;,\; \omega_{ij} = \begin{cases} 1 & \text{if } rt_{ij}(t) \geq 0 \\ 0 & \text{otherwise} \end{cases}$$

$$bid_i^{time}(t) = \overline{lp} + \left(\frac{b_i}{l_i} - \overline{lp}\right)\left(1 - \frac{\overline{rt}_i(t)}{rt_i^{max}}\right)^{\frac{1}{\beta}} \tag{2}$$

$$bid_i(t) = \alpha' bid_i^{resource}(t) + \beta' bid_i^{time}(t) \tag{3}$$

$0 \leq \alpha', \beta' \leq 1$

In these equations, $n_i^t$ is the number of remaining resources in the time t, which can be applied by applicant $t_i$, and $n_i^{max}$ is the maximum number of resources that might be applied by $t_i$. Remaining time of $t_i$ who is utilizing $r_j$ is calculated as $rt_{ij}(t) = d_i - st_j - l_i/cpu_j$. Let $rt_i^{max}$ be the maximum time of waiting for $t_i$. Different applicant's budget curve can be adjusted by changing $\alpha$ and $\beta$. In fig.4 different values of $\alpha$ is shown. It has the same shape for $\beta$ and σ (that will be introduced in next parts of paper).

After calculation of the budget of applicant, it is time to calculate the price of the resource. General policy of the resource owner is to service the applicant with the highest budget, in order to increase the utilizing the resource. In other words:

$$rp_j(t) = lp_j + (hp_j - lp_j)\left(\frac{st_j(t)}{wl_j(t)}\right)^{\frac{1}{\sigma}} \qquad (4)$$

Where, $rp_j(t)$ is the price of resource at time *t* and $st_j(t)$ is the current workload or the workload at the start time of the task at time *t*. $wl_j(t)$ is the workload of $r_j$ after the last allocation. In this equation, general trend is to decrease the price of the resource when the allocated resource is going to finish the task, in order to let other applicants to apply for it easier.

After the calculation of both budget and price, they are submitted to ResAg. It sorts the applicants' budgets in descent order and the resource prices in ascent order. Then, according to (5) final price is calculated as the average of the richest applicant ($bid_i^{max}$) and cheapest resource ($rp_j^{min}$):

$$fp(t) = \tfrac{1}{2}\left(bid_i^{max}(t) + rp_j^{min}(t)\right) \qquad (5)$$

Matrix *P* is then constructed according to the final prices. Resource allocation methods utilize this *P* for binding the resource to the applicant. But, an important issue is optimizing the values of matrix *P* to achieve more valuable goals like green computing.

This section was mostly adopted from [21]; and readers can refer to it for further details about the construction *P* and strategies of applicant and resource owners.

## 3    Optimization of Matrix *P* Based on the History of Latencies

### 3.1    Taking the Communication Latencies into Account

Definitely, there might be a geographical distance between the resource and applicant that causes latencies in communication. These latencies affect the whole performance of the system and increase the average response times. These negative results will be emerged while execution of fine-grained parallel programs. Trade-off between the latency factors that might be forked from the faults in the system of from the geographical distance and performance of system, encourages us for optimization of matrix *P* to achieve better performance. The main contribution is to maintain a history of latencies. The history of latencies can be taken into the account for constructing P. It will help to have better performance of system when there are similar resources. Then, ResAg can consider the latencies as a part of the weight of $fp$.

Construction of the table of latencies is done gradually. After each resource allocation, a record is added to the table of latencies that shows the latency between $t_i$ and $r_j$; or modifies the previous records between them. Let resource $r_j$ is allocated to applicant $t_i$ for the first time. Latencies of messages communication can be measured through sending acknowledge packets. Acknowledge packets might be more than once submitted from applicant to resource (and vice versa) to collect the average of the latencies:

$$\overline{LC}_{ij} = \sum_{q=1}^{p} PL_q$$

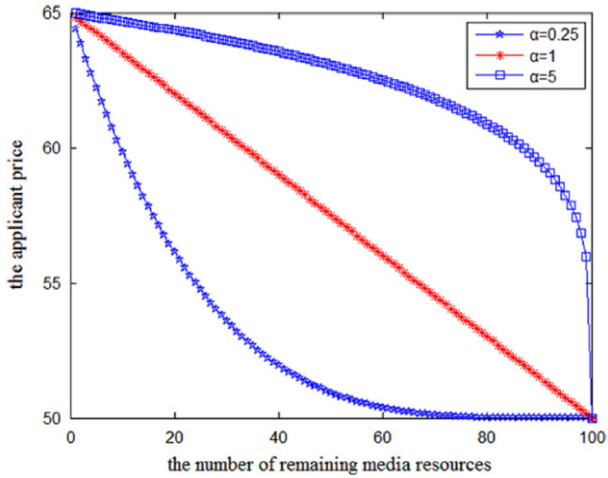

**Fig. 4.** Resource applicants' price considering remaining resources with different values of $\alpha$ [21]

Where $\overline{LC}_{ij}$ is the average latency of communication between applicant $t_i$ and resource $r_j$ and $PL_q$ is the latency of $q^{th}$ acknowledge message. Note that, there are no lower and upper bounds for $\overline{LC}_{ij}$ ($0 \leq \overline{LC}_{ij} \leq \infty$). Value *0* for the $\overline{LC}_{ij}$ is when the resource and applicant are located at the same node and there is no communication latency; and value $\infty$ is when the resource is faced with failure and it sends no response to the acknowledge message. So, it can't be utilized directly in the *P*. Here we need to change it to the scale of *0* to *1* in order to affect the *P* with the value of $\overline{LC}_{ij}$. Due to it, the average of all $\overline{LC}_{ij}$ that are stored in the specified table should be calculated as (6):

$$ALC(t) = \frac{1}{m \times n} \times \sum_{i=1}^{n} \sum_{j=1}^{m} \overline{LC}_{ij}$$
*if there is a record in table of latencies for $t_i$ and $r_j$* (6)

Where $ALC(t)$ is the average of latencies of communications between nodes $t_i$ and $r_j$ where they were allocated previously and there is a record in the table of latencies for them.

Equation (7) generates a number between *0* and *1*, for modification of matrix *P*:

$$TLC_{ij}(t) = 1 - \left(\frac{\overline{LC}_{ij}}{\overline{LC}_{ij} + ALC(t)}\right) \qquad (7)$$

Where $TLC_{ij}(t)$ is the impact of total latency between applicant $t_i$ and resource $r_j$ at the time *t*. $TLC_{ij}(t) = 0$ means that the resource is unavailable and it gave no response to the acknowledge packet ($\overline{LC}_{ij} = \infty$); accordingly, $TLC_{ij}(t) = 1$ means that there is no latency and resource and applicant are located at the same node ($\overline{LC}_{ij} = 0$).

Now, matrix LC can be constructed as (8):

$$LC = \begin{bmatrix} l_{11} & l_{12} & \cdots & l_{1n} \\ l_{21} & l_{22} & \cdots & l_{2n} \\ \vdots & \vdots & \vdots & \vdots \\ l_{m1} & l_{m2} & \cdots & l_{mn} \end{bmatrix} \quad s.t. \; 0 \leq l_{ij} \leq 1 \;,\; \sum_{i=1}^{m} l_{ij} = 1 \;,\; \sum_{j=1}^{n} l_{ij} = 1 \qquad (8)$$

Where $l_{ij}$ is the impact of latency between applicant $t_i$ and resource $r_j$ which is obtained from $TLC_{ij}(t)$.

### 3.2    Modification of Matrix *P* with LC

Now, it is time to have a consequent matrix at ResAg to do the resource allocation efficiently. It should be pointed out that in all the systems, policies regulate everything. So, some facilities to implement the policies should be considered. So, we will put a weight on the *P* and LC to be able to control them. The consequent is matrix *FP* as in (9):

$$FP(t) = \frac{1}{\theta + \lambda} \times (\theta P + \lambda LC) =$$
$$\frac{1}{\theta + \lambda} \times \begin{bmatrix} \theta p_{11} + \lambda l_{11} & \theta p_{12} + \lambda l_{12} & \cdots & \theta p_{1n} + \lambda l_{1n} \\ \theta p_{21} + \lambda l_{21} & \theta p_{22} + \lambda l_{22} & \cdots & \theta p_{2n} + \lambda l_{2n} \\ \vdots & \vdots & \vdots & \vdots \\ \theta p_{m1} + \lambda l_{m1} & \theta p_{m2} + \lambda l_{m2} & \cdots & \theta p_{mn} + \lambda l_{mn} \end{bmatrix} \qquad (9)$$

Where, $\theta$ is the weight of *P* and $\lambda$ is the weight of LC to implement the policies of system.

### 3.3    Side Issues

Inherently, the proposed method has some features to detect the unavailable resources. In fact, $\overline{LC}_{ij}$ can indicate the availability of resource $r_j$. It can be a good feature for handling the faults of system by isolating the crashed resources. This issue will also assist the migration strategies to have a better performance. Replication is not out of the circle, too.

The way of calculation of $\overline{LC}_{ij}$ for unavailable resources has a major drawback. When a node is unavailable, $\overline{LC}_{ij} \to \infty$; so, $FP_{ij}$ will be take a lower number, so that the resource $r_j$ never be allocated to any applicant task even after becoming available. This trap will practically omit the resource from the overlay network of resource-applicant graph. After repairing the failed resource, it won't be able to come back to the network. Due to solve this problem, some solutions must be taken into the account. For example, ResAg can set a timestamp for the resources that $\overline{LC}_{ij} = \infty$ to check their availability every time to time.

## 4    Simulation and Results

Implementation of proposed method and analyzing the results lead to better understanding about the efficiency of this method. In order to evaluate the performance of the proposed algorithm, we implement it by the CloudSim toolkit [23]. Each task is submitted according to Poisson distribution after its previous tasks, the length of each task is considered as a random number within [100000,200000], the number of tasks are considered between [100,1000], while the number of resources is between [30,50], the deadline $d_i$ of task $t_i$ is set according to (10), and the budget bi of task $t_i$ is set according to(11) [21].

$$d_i = st_j + random\left(\frac{l_i}{1.1 \times cpu_j}, \frac{l_i}{0.9 \times cpu_j}\right) \tag{10}$$

$$b_i = l_i \times random(0.9\overline{lp}, 1.1\overline{hp}) \tag{11}$$

Where $\overline{lp}$ and $\overline{hp}$ are the average values of the media resources' $lp$ and $hp$ [21].

Fig.5 shows a comparison between the response times of common method (which was talked in section 2) and the proposed method with optimization based on the communication latencies between the nodes. This indicates the efficiency of the LO method especially by increasing the number of tasks and passing more times.

In standard resource allocation, the common method is implemented, which the tasks come into the system and execute normally. In this way, latencies between the nodes are waivered. For example, when ResAg is allocating a resource to applicant, it does not consider the factor of latencies. So, it might choose a resource with highest latencies to be allocated to applicant. It will cause longer response times. So, the basic strategy is to let the ResAg to know about the latencies among the nodes of resources and applicants. Then, it can include the factor of latencies to make decisions about choosing the resources for allocation. This advantage improves the response times especially in the case of increasing the tasks. On the other hands, increasing the resources from similar types can also improve the performance of system in comparison with the common resource allocation. Because, it provides more choices for ResAg for selecting the best latencies.

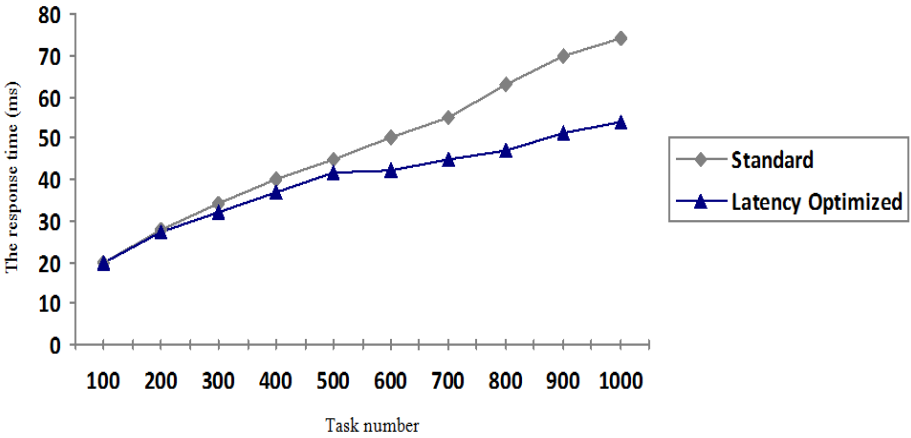

**Fig. 5.** Comparison of response time between the common and latency optimized methods

## 5     Discussion and Conclusion

The impact of communication latencies on total performance of Cloud systems encouraged us to optimize one of the most common resource allocation methods. Almost in all distributed systems, resource allocation is counted as a duty of Resource Agent. Both resources and applicants calculate their prices and budgets and send it to the ResAg. ResAg then makes decision to allocate the most appropriate resource to the best applicant. Due to it, ResAg constructs the matrix $P$ (as in section 2) based on the prices and budgets. Optimization of $P$ can lead to better performance. In our method, we considered a table in ResAg that holds the history of the resource allocation bindings with their average latencies. At first, this table has no record and after each allocation a record is added or updated. For next allocations, the values of latency impact will be taken into account for making decision. Accordingly, Matrix $FP$ is constructed where $FP_{ij}(t)$ is the possibility of resource $r_j$ to be allocated to applicant $t_i$. This value is partially obtained from the average latencies of previous allocations. Response time is considered as a metric for evaluation of current method. Results indicate the better response time rather than standard method, especially by increasing the number of tasks and passing time. Increasing the number of tasks and especially more resources from similar types, let the ResAg to have a more options for allocation. In this war, ResAg can select the resources with the best latencies as a part of their decisions. Besides, this method can be utilized to detect the failure of resources by measuring the latency of communications; so that the nodes with very high latencies are considered to be disconnected from the network. It is an important point for the other issues like migration, resource replication and fault-tolerance.